# Experimental Demonstration of DDoS Mitigation over a Quantum Key Distribution (QKD) Network Using Software Defined Networking (SDN)

E. Hugues-Salas[1], F. Ntavou[1], Y. Ou[1], J. E. Kennard[2], C. White[3], D. Gkounis[1], K. Nikolovgenis[1], G. Kanellos[1], C. Erven[2], A. Lord[3], R. Nejabati[1] and D. Simeonidou[1]

*(1) High Performance Networks group, (2) Centre for Quantum Photonics, School of Physics & Computer Science, Electrical and Electronic Engineering, and Engineering Maths, University of Bristol, BS8 1UB, UK (3) British Telecom (BT) Research and Innovation, Adastral Park, Ipswich, UK*

*e.huguessalas@bristol.ac.uk*

**Abstract:** We experimentally demonstrate, for the first time, DDoS mitigation of QKD-based networks utilizing a software defined network application. Successful quantum-secured link allocation is achieved after a DDoS attack based on real-time monitoring of quantum parameters.

Quantum key distribution (QKD) is considered an important cryptography method aimed at solving a number of security problems in communication networks [1]. In QKD, symmetric keys are generated by transmitting single photons from Alice to Bob over an optical channel. QKD enables detection of any eavesdropping attempt by Eve, based on the fundamental constraints of quantum mechanics.

In recent years, QKD technologies have achieved a significant level of maturity and successful field trials have been demonstrated using point-to-point QKD links to create key distribution networks [2]. However, it has been identified that QKD is vulnerable to Distributed Denial of Service (DDoS) attacks, typically aborting key establishment sessions whenever tampering or any strong perturbation is detected on the quantum channel, disrupting the key generation [3,4]. To overcome this problem, the provision of quantum-secured paths over a network configuration has been proposed [4]. In addition, the use of Software Defined Networking (SDN) is highly beneficial in quantum-secured optical networks since SDN adds flexibility and programmability together with a centralized management of optical resources [5]. To this end, we have demonstrated the use of SDN for time-sharing of QKD resources in which a programmable cost-effective network can be designed [6].

In this paper, we experimentally demonstrate, for the first time, an SDN-enabled circuit-switched optical network with QKD resources under DDoS attacks. An SDN application is developed to monitor the quantum parameters of the system, such as the secret key rate (SKR) and the quantum bit error rate (QBER). Based on this application, SDN can be used to detect link failures (due to attacks) and to provision newly secure paths to mitigate DDoS. In our testbed, we used three possible scenarios. The first scenario emulates a simple attacker directly over a standard single mode fiber (SSMF) link. The second addresses a multicore fiber (MCF)-based QKD network, which could be used in intra-data center applications to reduce the required dedicated SSMFs (i.e., expensive and unavailable). An attacker is applied in one of the adjacent cores of the MCF core that carries the encoded photon. Lastly, the third scenario represents a multi-hop (2-hops) link which can be found in meshed complex networks.

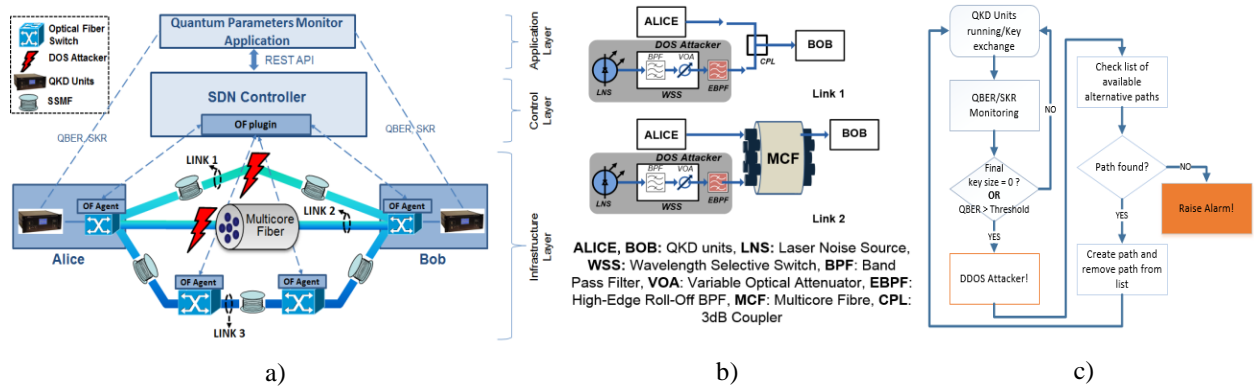

Fig. 1. a) SDN/QKD Optical Network testbed. b) DoS attacker for links 1 and 2. c) DDoS Mitigation Flowchart.

## 2. SDN-Enabled QKD Network Setup under DDoS Attack

Fig. 1a shows the SDN-enabled optical network with QKD resources. In the Alice node, a QKD-Alice unit (ID Quantique Clavis[2] ID3100) is connected to a low-insertion loss, large-port count optical switch (Polatis[TM]). Similarly, in the Bob node, a QKD-Bob unit is connected to a port of the optical switch. The optical switches are interconnected by three parallel optical links: *i)* the first link includes a 3dB coupler (Fig. 1b) in which the Raman noise from a tunable laser is filtered (0.7nm bandwidth) and a high attenuation is gradually reduced to add noise to the optical link at the 1552nm wavelength of the quantum channel; *ii)* in the second link, a 1km long, 7-core multicore fiber with an average core pitch of ~44.7μm and a loss of 0.2dB/km provides an alternative path between the QKD units. A DDoS attack is emulated by inducing crosstalk at 1552nm (Fig. 1b), in an adjacent core to the one carrying the QKD signal. *iii)* for the third secure path, a 520m long SSMF link is used to alleviate the attacks in the other two links. This last optical link passes through two intermediate nodes which are used to interconnect the Alice and Bob units. The total power loss of one of the 3 links is adjusted to be ~9dB, using a variable optical attenuator.

Fig. 1a also depicts the SDN architecture that is applied over the optical network, which continuously monitors the system's parameters and reacts in real time to the presence of a link failure. The Quantum Parameters Monitor (QPM) application monitors the QBER and the SKR from the QKD units and interfaces with the SDN Controller through the REST API. In the optical switches, an OpenFlow agent (OpenFlow 1.0) is installed to allow the connectivity to the SDN controller (OpenDaylight, Lithium). Fig. 1c shows the flow and decision making of the QPM application. More specifically, if the QKD units are not generating keys, i.e. the final key size is equal to zero or the QBER is above a specified threshold, the QPM application will detect this link failure and react to change the current optical path. The QPM application maintains a list of all three available paths in the order depicted in Fig. 1a, which have been pre-calculated between the pair of the QKD units, and selects the first path on the list. A reconfiguration decision will be made by the application and then sent to the SDN controller using an HTTP POST request. This decision contains the cross-connections to be performed for setting up the new optical path. Following this, the SDN Controller sends OpenFlow messages to the optical switches to create the suitable cross-connections. Once this process is undertaken, the key generation operation re-initializes through the new secure path and the monitoring procedure starts again to detect new attacks.

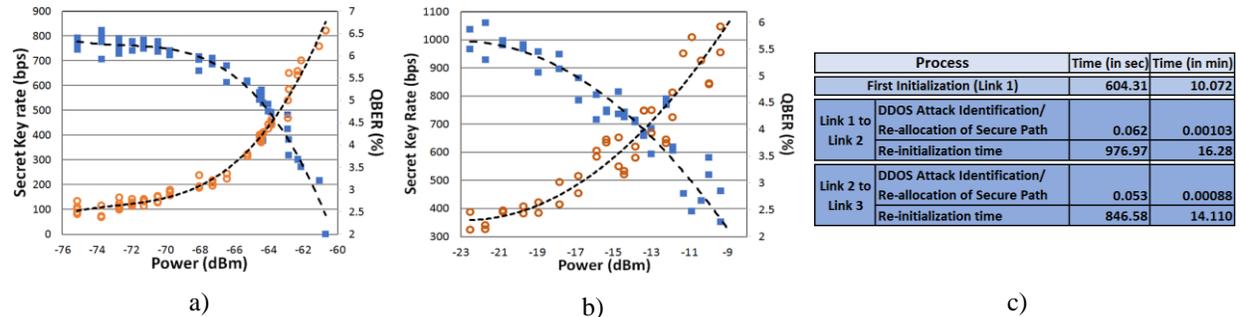

Fig. 2. a) SKR/QBER vs Power (DDoS Attacker) in link 1 (squares - left axis, circles - right axis). b) SKR/QBER vs Power (DDoS Attacker) over adjacent MCF core (squares - left axis, circles - right axis). c) Measured times during the DDoS attacks.

## 3. Experimental Results

To show the impact of the emulated DDoS attackers, in Fig. 2a-b the SKR and QBER curves are presented as a function of the optical power generated by the DDoS attackers in links 1 and 2. As it can be observed in Fig. 2a, the SKR and QBER are within the limits of 700b/s to 800b/s and 2.5% to 3%, respectively, up to a measured optical power of -68dBm (OSA Anritsu MS9710B, @0.07nm resolution, 70dB dynamic range). When the injected power in the 3dB coupler of link 1 is higher than -68dBm, the SKR (QBER) decreases (increases) until the generation of keys is not possible, since the QKD units detect an excessive number of errors unable to be corrected.

For the case of the MCF link, the crosstalk generated by injecting power in an adjacent core will impact the SKR and QBER above the measured power of -17dBm, as shown in Fig. 2b. The measured injected power level can be up to -9dBm before the system stops generating keys. It is important to mention that this power level depends on the number of cores used, where additional cores with optical power will set a lower power threshold due to the induced crosstalk by adjacent cores. Fig. 2c shows the measured times for different processes of the experiment. The times were measured for both the case the attacker is found in link 1 and the SDN controller changes to link 2, and for the case of link 2 being attacked and changing to link 3. It is observed that the time required by the SDN controller to identify the attack and setup the new optical path is negligible compared to the other processing times. The total time for the re-initialization after the change of the optical path and until a new key is generated is close to the first initialization time, which means that the impact of the switching is tolerably low. During the test, all the cross-connections required to transition between network configurations are switched simultaneously, which is time-efficient.

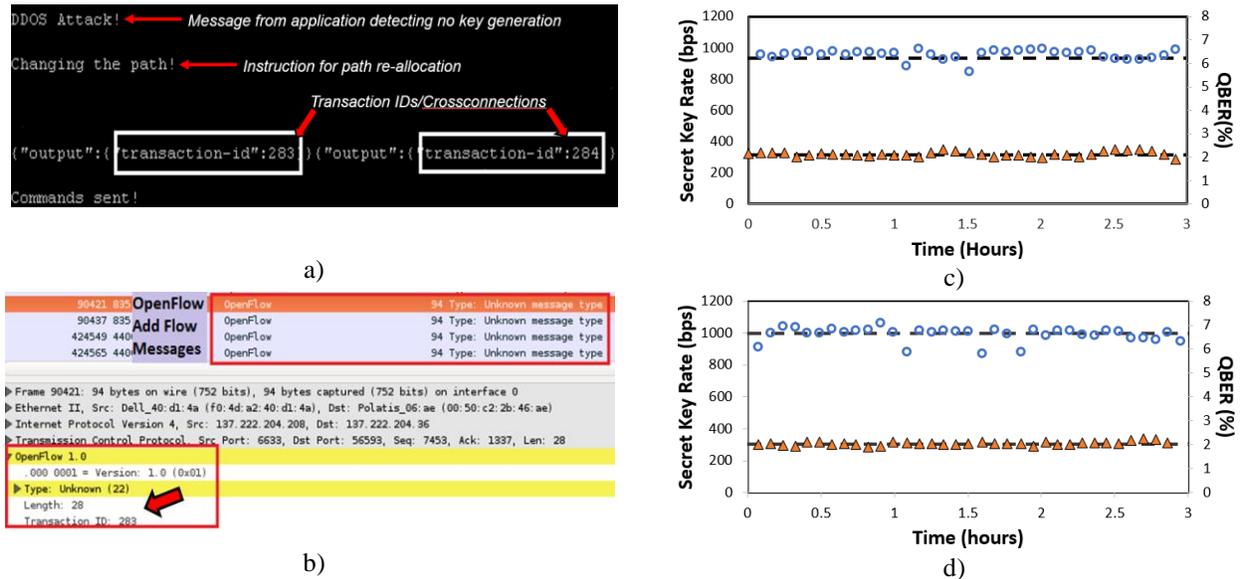

Fig. 3. a) Output of the QPM application algorithm. b) Wireshark captured OpenFlow messages. c) Measured SKR/QBER of link 2 following a DDoS attack event and d) Reference SKR/QBER on link 2 (triangle - right axis, circles - left axis).

Fig. 3a shows the output of the QPM application software after an event of DDoS attack, triggered by the absence of key generation. As shown in Fig. 3a, transaction IDs are created corresponding to OpenFlow messages sent from the SDN controller to the optical switch, as shown in the Wireshark screenshot of Fig. 3b for one transaction (ID 283). Each transaction ID corresponds to a cross-connection in the optical switch resulting in a new quantum-secured path. Fig. 3c illustrates the SKR and QBER measured over a period of 3-hours for link 2 after the DDoS attack occurred on link 1 and the quantum channel was switched to the new secured path. For reference, Fig. 3d shows the SKR and QBER of the same link 2 (MCF core) in the absence of DDoS attack in any link, meaning that switching from link 1 to link 2 wasn't required and link 2 was used as the QKD secure path. In both Figs. 3c-d, it is observed an average SKR and QBER of 950b/s and 2%, respectively, denoting again, that the switching to the new path undertaken by the SDN application does not impact the performance of the QKD system. We note that an in-field installation over much longer and more varied optical paths might not provide a choice between two paths of near-equivalent loss and noise, and therefore the re-routing of a path due to a DDoS attack could result in a change in performance.

## 4. Conclusions

A QKD network vulnerable to DDoS attacks was experimentally demonstrated. In this network, an SDN application was implemented for the mitigation of the attacks. This application monitored in real-time the quantum parameters of SKR and QBER and reacted in the event of lack of key generation, selecting an alternative route as a quantum channel. Three different link scenarios were used, proving continuous key generation and insignificant impact on the identification and re-allocation of secure path times compared to the QKD ones. This investigation shows the benefits of SDN applications in quantum networks for future secure communications.

## 4. Acknowledgements.

This work acknowledges EPSRC EP/M013472/1: UK Quantum Hub for Quantum Communications Technologies, EP/L020009/1: Towards Ultimate Convergence of All Networks and the EU Horizon 2020 METRO-HAUL project.

## 5. References

[1] H.K Lo, M Curty, and K Tamaki, "Secure quantum key distribution" Nature Photonics, **8**, 595-604 (2014).
[2] M. Sasaki, et. al., "Field test of quantum key distribution in the Tokyo QKD Network", Opt. Express, **19**, 10387-10409 (2011).
[3] P. Schartner and S. Rass, "Quantum key distribution and Denial-of-Service: Using strengthened classical cryptography as a fallback option", International Computer Symposium (ICS), Taiwan, (2010).
[4] R.Alléaume,et.al.,"Using quantum key distribution for cryptographic purposes: A survey", Theoretical Computer Science, **560,** 62–81 (2014).
[5] T. S. Humble and R. J. Sadlier, "Software-defined quantum communication systems". Optical Engineering. **53**. 086103-086103, (2014).
[6] A. Aguado, et. al., "Secure NFV Orchestration Over an SDN-Controlled Optical Network With Time-Shared Quantum Key Distribution Resources", Journal of Lightwave Technology, **35**, 1357-1362 (2017).